\documentclass[pre,amsmath]{revtex4-2}
\usepackage[pdftex]{graphicx} 
\usepackage{color}
\usepackage{bm}
\usepackage{epsfig}
\usepackage{latexsym}
\usepackage{nameref,hyperref}
\usepackage{xcolor}

\begin{document}
\title{Autotaxis of a Run-and-Tumble Agent via Reinforcement Learning}
\author{Ramesh Pramanik$^{1}$, Shradha Mishra$^2$ and Sakuntala Chatterjee$^1$}
\affiliation{(1) Department of Physics of Complex Systems, S.N. Bose National Centre for Basic Sciences, Salt Lake Sector 3, Kolkata 700106, India \\
(2) Department of Physics, Indian Institute of Technology (BHU) Varanasi 221005, India}
\begin{abstract}
We study the run-tumble motion of a single autochemotactic agent that secretes and senses chemical cues using a reinforcement learning (RL) framework. In a chemical environment that changes dynamically via secretion, diffusion, and degradation of attractant chemicals, we study the effectiveness of RL for a successful navigation strategy. We show that RL enhances chemotactic performance. Starting from a random initial position, the probability of finding the agent in the favorable region (near the attractant peak) increases faster with time and reaches a higher value in the long time limit when RL is used, compared to the case when the agent performs usual run-tumble chemotaxis without using RL. We also characterize the chemotactic performance of the agent by varying different parameters like rates of secretion, degradation, and diffusion. Importantly, we observe a non-monotonic dependence of the chemotactic performance on the learning and exploration rates, indicating the existence of an optimal RL strategy for the most efficient navigation. We explain these observations as a result of the interplay between the pre-existing attractant environment and the agent-generated attractant profile. 
\end{abstract}

\maketitle
\newcommand{\be}{\begin{equation}}
\newcommand{\ee}{\end{equation}}
\newcommand{\bea}{\begin{eqnarray}}
\newcommand{\eea}{\end{eqnarray}}

\section{INTRODUCTION}

Motion of certain organisms in response to chemical signals received from their environment is called chemotaxis \cite{eisenbachbook, berg2008coli, Moore}. Typical examples include bacteria searching for nutrients or avoiding toxic substances \cite{Young}, immune cells moving toward infected regions \cite{Vargas}, and sperm cells swimming toward the egg \cite{Friedrich, Hamdoun, Kaupp, Eisenbach2025, Lange}. In certain situations, apart from following the chemical cue received from the environment, the moving organism can also secrete its own chemical, which then acts as an attractant. This phenomenon is called autochemotaxis or autotaxis \cite{PhysRevLett.95.128103,PhysRevE.80.031122,Mokhtari}. A well-known example of self-generated chemotactic signaling is found in Dictyostelium discoideum, where starving amoebae secrete and sense cAMP, generating propagating chemical waves that coordinate collective migration and aggregation into multicellular structures \cite{Konijn,Gregor}. Auto-chemotactic signaling also plays an important role in immune responses and wound healing, where neutrophils amplify chemotactic signals through the secretion of leukotriene B4 (LTB4), thereby recruiting additional immune cells to sites of inflammation \cite{Afonso}. Self-generated chemotactic signaling can also drive pattern formation in bacterial populations. For example, Escherichia coli cells secrete and respond to attractants, leading to collective aggregation and the spontaneous emergence of ordered spatial patterns such as spots and stripes \cite{Budrene}. Thus in a wide variety of biological systems autotaxis gives rise to highly interesting effects like self-trapping, aggregation, and pattern formation which have been studied both experimentally and theoretically \cite{Tsori_2004, PhysRevLett.131.088201, Amar2016, PhysRevE.98.052411, Livne, Mukherjee, Ma2016}.

To theoretically model such phenomena by directly solving coupled time-evolution equations is computationally challenging because of the multiple time scales associated with various processes like diffusion and degradation of the chemical, secretion by the organism, intracellular reactions in the signaling network of the organism, and motion of the organism. In this context, reinforcement learning (RL) provides a simple and efficient alternative framework. Instead of explicitly solving the coupled time-evolution equations for the chemical field and the cell position, an RL algorithm considers an agent that evolves through experience by minimizing a cost or maximizing a reward \cite{sutton2018reinforcement,panait2005cooperative, busoniu2008comprehensive, kubat2017introduction, watkins1992q, NIPS2005_9752d873,kanai, NIPS2005_9752d873, Mohan, schulman2017}. This learning-based formulation naturally captures feedback, adaptation, and decision-making without requiring detailed knowledge of microscopic mechanisms, making RL algorithms computationally efficient. Thus, RL offers a flexible and scalable approach to explore how a cell can learn to navigate, localize, or escape using autochemotactic signals.

Initially formulated in the context of artificial intelligence, reinforcement learning has found broad applications in physical and biological systems \cite{sampat2022ordering, Kumar_2025, Neftci2019, Treloar}, stochastic control \cite{Yongacoglu, Zhu, MunosBourgine1997}, cellular decision-making \cite{TEJEDOR2020101836, PhysRevX.9.031012}, and collective motion \cite{durve2020learning, Hou2019}. In particular, reinforcement learning has been recently applied to chemotaxis problems, where the agent learns to navigate chemical landscapes through adaptive responses \cite{PhysRevE.111.014106, D5SM01040C}. These studies show that there exists an optimal combination of the learning rate $\alpha$ and the exploration parameter $\epsilon$ that yields the most effective chemotactic response. At this optimum, the agent balances directed movement toward chemical gradients with sufficient exploration to avoid getting trapped in local concentration maxima of attractant chemical \cite{PhysRevE.111.014106}. 

In this work, we extend the RL framework developed in our previous study \cite{PhysRevE.111.014106} for run-and-tumble chemotaxis in externally prescribed chemical landscapes to a single auto-chemotactic E. coli cell that creates and senses its own chemical field. We ask the question whether RL improves the chemotactic performance in a complex and time-varying chemical landscape. We explicitly compare the long time performance of our RL agent and a non-learning agent to show that RL indeed yields better performance. To navigate in a chemical environment, an {\sl E.coli} cell uses its run-tumble motion, which consists of alternating between straight `runs' and random reorientation events, or `tumbles' \cite{Berg1972, PhysRevE.99.012402}. In a medium where the concentration of attractant is the same everywhere, the switching between run and tumble modes occurs at a constant rate. But in the presence of an attractant concentration gradient, the cell modulates its tumbling rate to prolong runs in the favorable direction ({\sl i.e.} up the gradient) and shorten the ones in the opposite direction (down the gradient) \cite{Gennes, kafri, ladimirov, PhysRevE.97.032420, PhysRevE.105.014411, Mandal_2025, Chatterjee_2022, Pramanik2025, Mandal2022, PhysRevE.103.L030401}. This mechanism allows the cell to climb up the gradient and localize itself in a region with high attractant concentration. Notably, the small size of an {\sl E.coli} cell prohibits it from reading off the spatial gradient directly \cite{Mingshan}. Instead, the cell relies on temporal integration. By comparing the attractant concentration experienced in the recent past and in the distant past, the cell distinguishes between a favorable direction and an unfavorable direction \cite{Tedesco,PhysRevLett.100.238101,PhysRevE.91.042714, Segall, Block1, Brown, Macnab}. In theoretical modeling a non-Markovian random walker with run-tumble motility is often used to describe {\sl E.coli} chemotaxis \cite{Gennes,kafri,PhysRevLett.100.238101}. The presence of memory in these systems makes them analytically and computationally challenging to deal with.

We model the  {\sl E.coli} cell as an RL agent and use an $\epsilon$-greedy learning strategy \cite{sutton2018reinforcement}. In this framework, the agent stores information about past states and their outcomes in a $Q$-matrix, which guides the decision-making process \cite{watkins1992q}. Unlike conventional reinforcement-learning approaches, where the agent is trained until a converged policy is obtained and subsequently exploits this policy \cite{Putzke2023, Monderkamp_2022, durve2020learning}, here we employ an online learning strategy in which learning and navigation occur simultaneously. The agent continuously updates its Q-values based on its current interaction with the environment and selects its action using the presently learned Q-values. Thus, the learning rate and exploration parameters directly influence the agent's behavior during navigation rather than merely controlling convergence toward a final policy. This approach is particularly relevant for the present problem, where the chemical environment is dynamically evolving and is generated by the agent itself. Our aim is therefore not to obtain a globally optimal converged policy, but to investigate how continuous learning can enhance navigation in a dynamically changing, self-generated environment. The agent follows the chemical gradient using its run-tumble motion. It also modifies the chemical environment by secreting chemicals along its trajectory. In addition, the chemical molecules undergo diffusion and degradation in the medium. We are interested in the chemotactic performance of the agent in the long time limit. This performance actually reflects how successful the RL strategy is. We quantitatively compare the performance of the RL agent with that of an {\sl E.coli} cell modeled as a run-tumble random walker with memory, without using RL \cite{kafri,PhysRevE.91.042714,PhysRevLett.100.238101}. We find that not only the RL agent performs better in the long time limit, but it also improves its performance faster with time. Thus RL offers a successful and efficient strategy for navigation in a dynamical chemical environment.

The RL agent responds to the attractant trail created by itself, as well as to the attractant gradient already present in the medium, which may have been generated by other autotactic agents or it may even have been set up in a controlled environment in a lab. The resulting behavior of the RL agent is often a result of competition between these two gradients. At short times, the agent attempts to move towards the local attractant peak and localize there. But as it spends more time near the peak, due to its own secretion, the peak becomes more and more pronounced, trapping the agent for a long time at that location. Thus, the shape of the initial attractant profile plays an important role here. We mainly consider three different types of initial attractant concentration: (i) a spatially uniform profile, (ii) a profile with multiple peaks of equal height, and (iii) a profile with multiple peaks of unequal heights. We study how the long-time performance of the agent depends on the different parameters and whether there is an optimal set of parameters that yields the most efficient chemotaxis. In all cases we find the performance increases with secretion rate and decreases with degradation rate. Interestingly, the performance shows a peak as the chemical diffusivity is varied. Apart from the long-time performance, we also measure the search time, defined as the mean first passage time of the agent at the attractant peak. This is particularly relevant when the chemical landscape features multiple peaks of unequal heights, and the agent needs to escape the local peaks and find the global peak. We find the most efficient search strategy involves an optimum balance between learning and exploration by the RL agent.

In Sec. \ref{sec:model}, we describe the model in detail and explain the RL algorithm. In Sec. \ref{sec:RL}, we present our results for the RL agent. We compare the performance of the RL agent with another one without using RL in Sec. \ref{sec:nmrw}. Some concluding remarks are included in Sec. \ref{sec:con}.

\section{FORMULATION OF RL ALGORITHM}
\label{sec:model}

\subsection{$Q$-learning: state, action and cost}

We consider a reinforcement learning agent executing run-and-tumble motion in a one-dimensional system of length $L$ with periodic boundary conditions. At time $t=0$ an attractant concentration profile $[L](x,0)$ is set up in the system. The chemical concentration profile subsequently evolves in time, as described later in this section. The state of the agent is denoted by $S$. The RL agent can be in either of the two states, `high' or `low', based on the local attractant concentration. If the attractant concentration at the agent’s position is lower than the average attractant level, then the agent is in the low state; otherwise, it is in the high state. Similar state representations based on averaged or integrated observables have been employed in a variety of reinforcement-learning studies, including the autonomous discovery of critical parameters in the Ising model \cite{Man2025}, motor exploration \cite{Zai2026}, noisy chemotaxis \cite{Alonso2024}, and reaction-channel identification \cite{Liang_2023}. The action $A$ of the agent is also two-fold. During its run-and-tumble motion, the agent can either run in the current direction (leftward or rightward) with a fixed velocity $v$ or it can reverse its direction (tumble). These two actions are also referred as `persist' and `reverse', respectively. 

\begin{figure}[hbt!]
\centering
\includegraphics[width=0.95\linewidth]{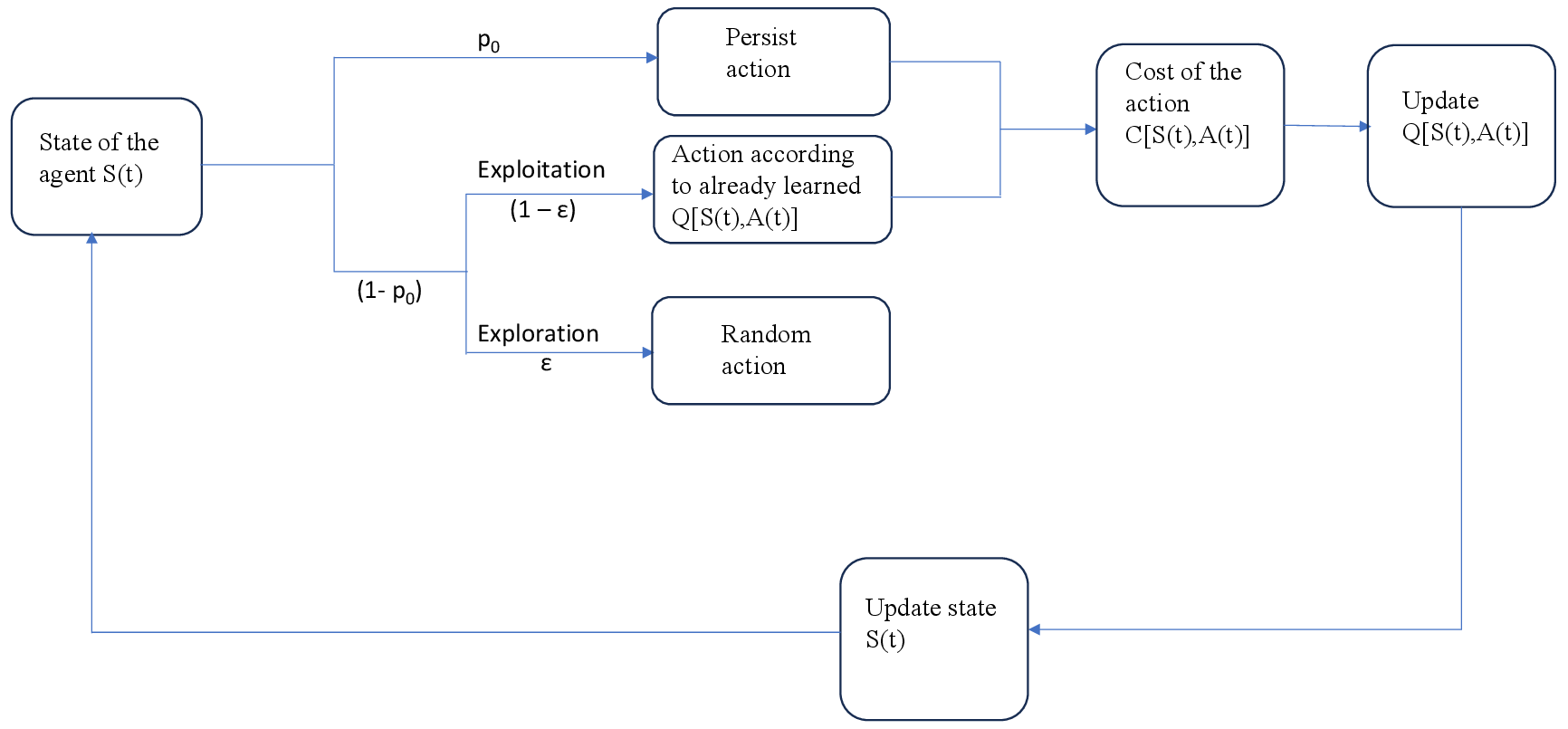}
\caption{Block diagram showing different stages of RL algorithm used here.}
\label{FIG:Flowchart_RL}
\end{figure}

To model how past experience in the chemical environment influences the agent’s motility, we describe its decision-making within a reinforcement learning framework based on $Q$-learning. In this framework, the agent’s past learning experiences are stored in a $Q$-matrix whose rows represent the state $S$ and columns stand for action $A$. In our case the $Q$-matrix is therefore $2 \times 2$ dimensional. At each time step, the agent evaluates the cost $c[S,A]$ associated with the chosen action and updates the corresponding $Q$-matrix element according to the $Q$-learning rule \cite{watkins1992q,Jia},
\begin{equation}
Q_{t+1}[S,A] \leftarrow Q_t[S,A] + \alpha \left( c[S,A] - Q_t[S,A] \right),
\end{equation}
where $\alpha$ is the learning rate. The $Q$-matrix is initialized with all elements set to zero at $t=0$. The agent follows an $\varepsilon$-greedy strategy: with probability $1 - \varepsilon$, it exploits past experience by selecting the action associated with the lowest $Q$ matrix element, while with probability $\varepsilon$, it explores by choosing an action at random, independent of the current $Q$-estimates. A schematic diagram of the RL algorithm is shown in Fig.~\ref{FIG:Flowchart_RL}.

Within the $Q$-learning framework described above, the cost $c[S,A]$ controls the learning of the agent. The assignment of $c[S,A]$ follows the principles of reinforcement learning and provides the physical link between the agent’s motility and its interaction with the chemical environment. Since we are interested in designing an RL algorithm that can successfully capture the behavior of {\sl E.coli} in an attractant gradient, we assign cost to execute a run or a tumble based on the actual behavior shown by {\sl E.coli}. Below we briefly describe the run-tumble motion of {\sl E.coli} and explain how we use input from here to define $c[S,A]$ in our model. 

\subsection{Run-tumble motion of  {\sl E.coli}}

\begin{figure}[hbt!]
\centering
\includegraphics[width=0.6\linewidth]{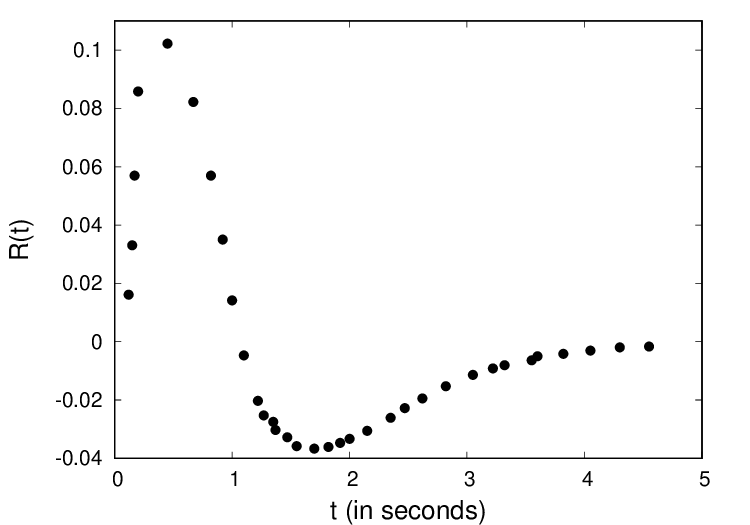}
\caption{The bilobe response kernel $R(t)$ of a wild-type {\sl E.coli} cell shows a sharp positive peak and a shallow negative peak such that the total area under the curve is zero. This curve is constructed from experimental data in \cite{Segall}. Instead of using the complete dataset, a few representative points are used for computational simplicity.}
\label{FIG:LANDSCAPE}
\end{figure}

During its run-tumble motion, an {\sl E.coli} cell performs directed runs for a certain length of time, followed by an abrupt change in direction when it tumbles. In a homogeneous attractant environment, the typical run duration of a wild-type {\sl E.coli} cell is $\tau \simeq 1s$. Tumbling events are almost instantaneous and occur with a small rate $1/ \tau$. Such run-tumble motions are active motion for which the tumbling rate always remains small. In the presence of an attractant gradient in its environment, the cell modulates its tumbling rate. By comparing the attractant concentration along its trajectory in the recent past and in the distant past, the cell reads off the gradient. This comparison is done with the help of a memory kernel, which has been measured experimentally \cite{Segall}. In Fig. \ref{FIG:LANDSCAPE} we show this memory kernel, which has a characteristic bilobed structure, with a positive peak at $t \simeq 0.5\,\mathrm{s}$ followed by a shallower negative lobe near $t \simeq 1.5\,\mathrm{s}$, and decays to zero for $t \gtrsim 4\,\mathrm{s}$. The tumbling rate can be written as 
\be 
\frac{1}{\tau}\left ( 1- a \int_{-\infty}^ t dt' R(t-t') [L](t') \right ) \label{eq:memory}
\ee
where $[L](t')$ is the attractant concentration experienced by the cell at a past time $t'$. The parameter $a \ll 1$ ensures that the magnitude of the second term remains significantly smaller than $1$. Throughout this study we have used $a=0.1$, but any other choice will not change our conclusions. When the cell is climbing up the gradient, the attractant level in the recent past is higher than that in the distant past. The tumbling rate then decreases, and the run gets longer. Similarly, when the cell is going downhill, the attractant level experienced in the recent past is lower than the distant past, in which case the tumbling rate increases, allowing the cell to tumble quickly and choose a new run direction. The downhill runs are therefore shortened. Longer uphill runs and shorter downhill runs cause a net drift motion of the cell up the gradient.

\subsection{Run and tumble by RL agent } \label{sec:rnt}
To assign cost for the two actions, run and tumble, by the RL agent, we follow the similar principle shown by an {\sl E.coli} cell. As described in the previous subsection, the attractant gradient along the past trajectory influences the tumbling rate. Therefore, for a given state $S$ of the agent, the cost for tumble is calculated based on past trajectory, as in Eq. \ref{eq:memory},  
\begin{equation}
c[S,tumble] = \int_{-\infty}^t dt' \, R(t - t') \, [L](t').
\label{eq:ONE}
\end{equation}
In contrast, the cost of persisting in the current direction $c[S, run]$ is set to zero. When the agent moves up a concentration gradient, $c[S,tumble]$ is positive, making tumbling more costly than running and thus favoring the run. Conversely, when moving down the gradient, tumbling cost becomes negative, increasing the likelihood of tumbling. Note that even when tumbling probability increases, it still remains quite small, such that at any time the agent is much more likely to persist in its current direction than tumble. In our model at every time step $\delta t$ we allow the agent to persist with a high probability $p_0$, and with the remaining probability, $(1 - p_0) \ll 1$, the agent invokes a reinforcement learning decision process to choose whether to persist or reverse its direction. In our study, we set $p_0 = 0.98$ and $\delta t = 0.01s$, which yields an average run duration of approximately $1s$ (corresponding to $100$ discrete time steps), consistent with observed run lengths in \textit{E. coli} motility \cite{Pramanik2025}. Also, motivated by the typical run speed of \textit{E. coli}, we use run speed $v = 20\,\mu\mathrm{m/s}$ \cite{berg2008coli}. Therefore, the displacement during one time step $\delta t$ is $v\delta t = 0.2\,\mu\mathrm{m}$, which is taken as the unit lattice spacing.

\subsection{Secretion, diffusion, and degradation of attractant}

Apart from the RL agent, the attractant profile also has its own dynamics. The agent releases attractant molecules at the rate $\gamma$. In our simulations, in a time step $\delta t$ we add one attractant molecule with a probability $\gamma \delta t$ at the site where the agent is present. To implement degradation, we remove one molecule on average from all sites where attractant is present. This process is performed with probability $\beta \delta t$ for each site. The chemical molecules also show diffusive motion on the lattice. If two neighboring sites have an unequal number of molecules, then one molecule hops from the higher occupancy site to the lower occupancy site, with probability $D \delta t$. In Table \ref{tab:params} we list the values of all these rate parameters and other simulation parameters.

\begin{table}[h!]
\caption{Model parameters used in the simulation.}
\centering
\begin{tabular}{|c|c|c|}
\hline
\textbf{Parameter} & \textbf{Description} & \textbf{Value / Range} \\
\hline
$L$ & System size & $1000~\mu\text{m}$ \\
\hline
$\gamma$ & Chemical deposition rate & $0$–$50~\text{s}^{-1}$ \\
\hline
$\beta$ & degradation rate per site &  $0$–$0.1~\text{s}^{-1}$ \\
\hline
$D$ & Hopping rate per site & $0$–$60 ~\text{s}^{-1} $ \\
\hline
$\alpha$ & Learning parameter & $0$-$0.4$ \\
\hline
$\epsilon$ & Exploration parameter & $0$-$0.8$ \\
\hline
$v$ & Agent swim speed & $20~\mu\text{m/s}$ \\
\hline
$\delta t$ & Time step & $0.01~\text{s}$ \\
\hline
$p_0$ & Persistence probability of agent & $0.98$ \\
\hline
$[L]_0$ & Initial mean number of attractant molecules & $20 $ \\
\hline

\end{tabular}
\label{tab:params}
\end{table}


\section{Behavior of RL Agent in Different Attractant Environments}
\label{sec:RL}

Using the above RL algorithm, we monitor the behavior of the agent in an attractant environment $[L](x,t)$, which changes with time either via diffusion, evaporation, or active deposition by the agent. After the agent is introduced in the medium, it attempts to climb up the attractant concentration gradient, creating a chemical trail along its way. If the agent finds a local attractant peak, it tends to localize there and, by depositing more attractant at that position, makes the peak even stronger. Thus the behavior of the agent as well as the shape of the attractant profile depends on the choice of the initial profile $[L](x,0)$. We consider three types of initial profiles: a spatially uniform profile, a profile with multiple peaks of equal height, and a profile with multiple peaks of unequal heights. The motivation for considering different initial concentration profiles arises from the fact that the chemical environment experienced by an agent depends on its history and the prior activity of other agents. When an agent enters the medium for the first time, it encounters an almost spatially uniform concentration field $[L](x,0)$. However, if other agents have previously traversed the medium, the concentration field becomes nonuniform due to earlier chemical secretions. In such cases, $[L](x,0)$ can exhibit significant spatial variations, with regions of higher or lower concentration depending on the history of agent activity. To capture these environmental heterogeneities, we consider simple model concentration profiles with sinusoidal modulations. Specifically, we employ a single sinusoidal profile and a superposition of two sine waves to represent varying degrees of spatial complexity. More realistic concentration landscapes can be generated by superposing multiple sinusoidal components, encompassing a broad class of spatially varying attractant environments. Our results are not limited to these specific choices and remain valid for a broad class of initial concentration profiles characterized by either equal-height or unequal-height peaks. We have considered the sinusoidal chemical profile and the profile with multiple peaks of different heights in order to study the competition between the externally imposed chemical field and the self-generated chemical field produced by the agent. In the sinusoidal case, the externally imposed chemical hills tend to guide the agent toward the concentration maxima, and we investigate how efficiently the RL dynamics enables the agent to localize near these peaks. In the case of multiple hills with different heights, the landscape contains both local and global maxima, allowing us to study how efficiently the agent can identify and reach the global attractant peak in a complex inhomogeneous environment. We consider a one-dimensional system of length $L$ with a periodic boundary condition. We monitor the chemotactic performance of the agent by changing different model parameters like deposition rate $\gamma$, chemical hopping rate $D$, degradation rate $\beta$, exploration rate $\epsilon$, and learning rate $\alpha$.

\subsection{Homogeneous Attractant Environment}

\begin{figure}[hbt!]
\centering
\includegraphics[width=0.45\linewidth]{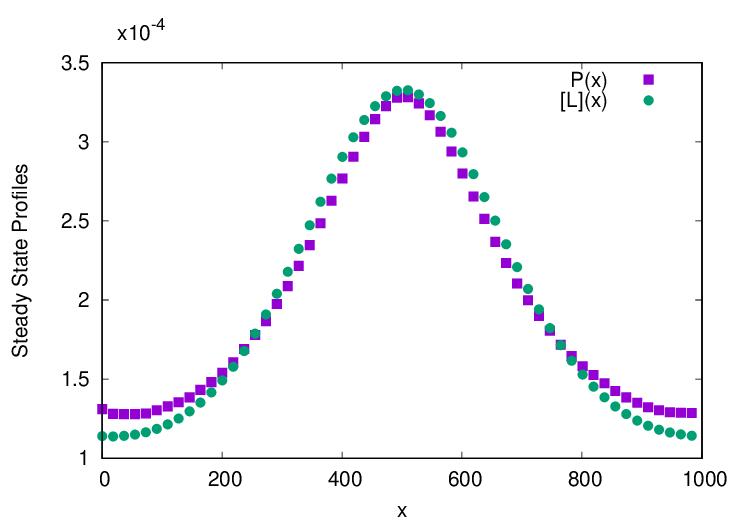}
\caption{Steady state position distribution of the agent along with the attractant concentration profile (normalized). Here the spatial coordinate $x$ is in micron ($\mu$m). The parameters are $D = 20 s^{-1}$, $\gamma = 50 s^{-1}$, $\beta = 0.005 s^{-1}$, $\alpha = 0.1$, and $\epsilon = 0.1$.} \label{FIG:FLAT_CX_PX}
\end{figure}

\begin{figure}[hbt!]
    \centering
    \includegraphics[width=0.45\linewidth]{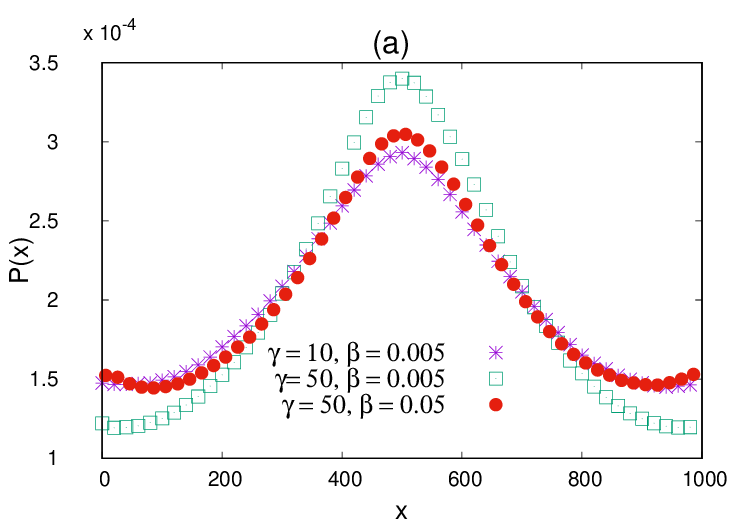}
    \includegraphics[width=0.45\linewidth]{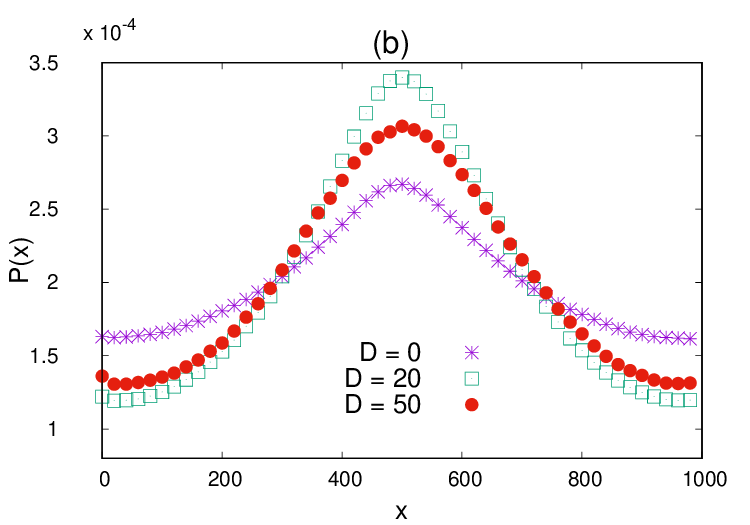}
    \includegraphics[width=0.45\linewidth]{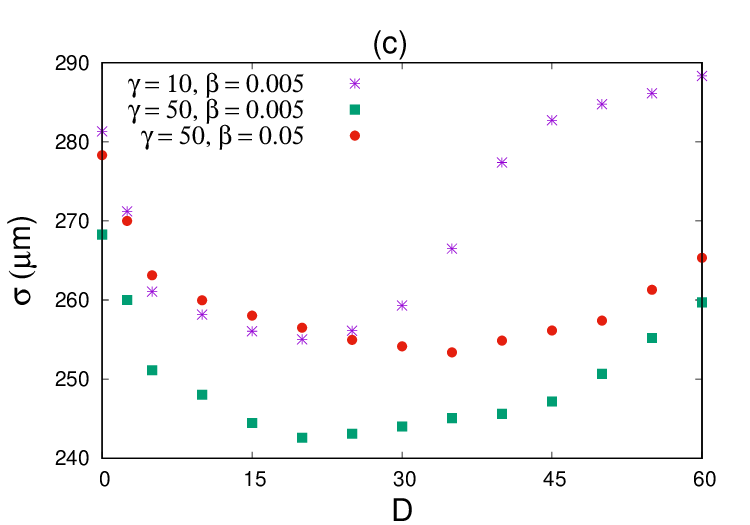}
    \includegraphics[width=0.45\linewidth]{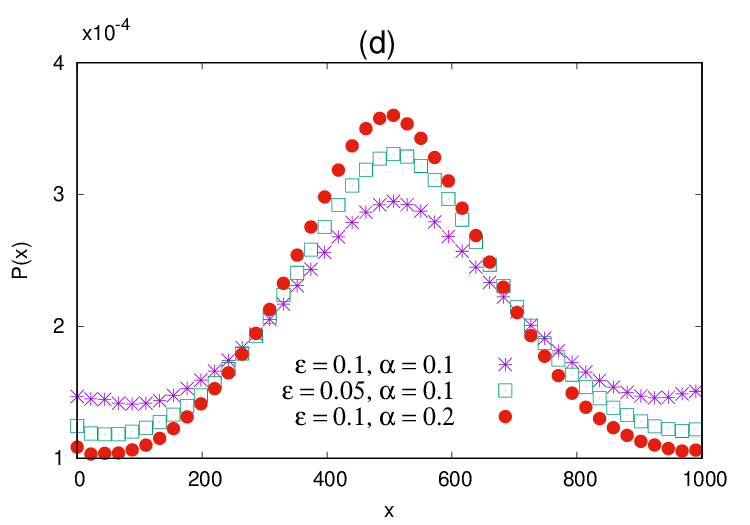}
    \caption{(a) Steady state position distribution $P(x)$ for different deposition rates $\gamma$ and degradation rates $\beta$ in the presence of an initially homogeneous attractant profile. Increasing $\gamma$ enhances localization of the agent around its initial position $x_0 = L/2$, while increasing $\beta$ weakens localization. We have used $D = 20 s^{-1}$, $\alpha = 0.1$, and $\epsilon = 0.1$. (b) Steady-state $P(x)$ for various $D$. Localization at $x_0$ exhibits a nonmonotonic dependence on $D$. Here $\gamma = 50 s^{-1}$, $\beta = 0.005 s^{-1}$, $\alpha = 0.1$, and $\epsilon = 0.1$. (c) Steady-state standard deviation $\sigma$ of the position distribution as a function of $D$. Here $\sigma$ shows a clear minimum at intermediate $D$, indicating optimal localization. Here $\alpha = 0.1$ and $\epsilon = 0.1$. (d) Steady-state position distribution $P(x)$ for different values of the learning rate $\alpha$ and the exploration rate $\epsilon$. The probability of finding the agent at its initial position $x_0 = L/2$ increases with $\alpha$ and decreases with $\epsilon$. The parameters are $\gamma = 50 s^{-1}$, $\beta = 0.005 s^{-1}$, and $D = 50 s^{-1}$. The spatial coordinate $x$ is measured in micron ($\mu$m), and $D$ is in $s^{-1}$.}
  \label{FIG:Homo_chemical_field}
\end{figure}

We start with an initially flat attractant profile and the RL agent placed at a fixed initial position $x=L/2$. The position distribution $P(x,t)$ of the RL agent starts from the delta function $\delta (x-L/2)$ at $t=0$ and begins spreading as time goes on, finally reaching a steady state in the long time limit. We evolve the system for $10^{6}$ time steps to ensure that the steady state is reached. The attractant profile $[L](x,t)$ is flat initially, but because of deposition from the agent, a peak develops around the agent position at $x=L/2$. As the agent moves and the chemical diffuses, this peak spreads in the medium. Finally, in the long time limit, the attractant profile and position distribution show similar spatial dependence, which ensures a successful chemotactic strategy. In Fig. \ref{FIG:FLAT_CX_PX} we compare the steady-state behavior of agent position distribution and attractant concentration and verify this. To investigate how chemotactic performance depends on deposition rate $\gamma$ and degradation rate $\beta$, we plot steady-state position distribution $P(x)$ for three sets of $\gamma$ and $\beta$ values in Fig. \ref{FIG:Homo_chemical_field}a. By looking at the width of $P(x)$, one can quantify performance. A small width means the agent is more strongly localized near the attractant peak, which indicates a good performance. Our data in Fig. \ref{FIG:Homo_chemical_field}a show that localization increases with $\gamma$ and decreases with $\beta$. A large deposition rate helps create a sharp attractant peak at $x=L/2$, and whenever the agent tries to move away from this zone, the cost for tumble drops sharply, causing the agent to reverse its direction and move towards the peak again. On the other hand, a large degradation rate erodes the chemical trail created by the agent and allows $P(x)$ to spread more widely in the medium. 

Interestingly, the chemical diffusion affects the agent performance in a non-monotonic way. In Fig. \ref{FIG:Homo_chemical_field}b we plot $P(x)$ for three different values of $D$ and observe that with increasing $D$ values, $P(x)$ first becomes narrower and then wider. In Fig. \ref{FIG:Homo_chemical_field}c we directly plot the standard deviation $\sigma$ of $P(x)$ as a function of $D$ and find a minimum. For different values of $\gamma$ and $\beta$, the position of the minimum changes, and $\sigma$ also takes different values, but there always exists an optimum $D$ at which the agent shows strongest localization. This can be explained in the following manner. When $D$ is very small, the attractant deposited by the agent cannot diffuse far. So the attractant peak created by the agent around $x=L/2$ remains confined in the same region even after a long time. Even if the agent ventures slightly away from this zone, it does not feel the presence of the peak anymore and behaves as if it is in a flat medium. On the other hand, for intermediate $D$, the attractant chemical diffuses to a larger distance such that a gradient is felt even far away from the peak. This helps the agent to relocate near the peak. For very large $D$, however, the attractant gradient becomes very weak, and the agent experiences an almost homogeneous environment. $P(x)$ becomes wider again. 

Finally, in Fig.~\ref{FIG:Homo_chemical_field}(d) we present the steady-state position distribution $P(x)$ of the agent for different values of the learning rate $\alpha$ and exploration rate $\epsilon$. We find the localization decreases with $\epsilon$ and increases with $\alpha$. As the exploration rate $\epsilon$ increases, the agent becomes more likely to take random actions, enabling broader exploration of its environment, thereby decreasing localization. Therefore in the limit $\epsilon =0$ strongest localization is obtained. On the other hand, for a fixed $\epsilon$ as $\alpha$ is increased, the agent is able to rapidly update the $Q$ matrix in response to changes in its chemical environment. This allows the agent to reinforce trajectories that lead to favorable outcomes, {\sl i.e.} spending a long time in regions of higher self-deposited attractant, thereby enhancing localization around the initial position. As mentioned in Sec. \ref{sec:rnt}, we have throughout considered $p_0=0.98$ here. But we have verified (data not shown here) that choosing a small value of $p_0$ decreases effective run length and keeps the agent confined near its initial position for longer time.

\subsection{Attractant Peaks of Equal Heights}

The behavior of the agent is significantly different when the initial attractant profile $[L](x,0)$ has spatial variation. In this subsection we consider an attractant profile which has multiple peaks of equal height. We use a simple sine function to model such a profile, $[L](x,0) = [L]_0 +[L]_a \sin\left( \frac{2\pi x}{\lambda} \right)$. We choose $\lambda=L/2$ such that there are two attractant peaks in our system. To quantify chemotactic efficiency of the agent in this case we consider a contrast measure, defined as $\eta = \frac{P_{1} - P_{2}}{P_{1} + P_{2}}$, where $P_{1}$ and $P_{2}$ denote the steady state probabilities of finding the agent near the attractant maximum and minimum, respectively. For the chemical profile with multiple peaks of different heights, the maximum and minimum concentration regions are indicated by the blue and red shaded regions, respectively, as shown in Fig.~\ref{FIG:2SIN}(a). A higher value of $\eta$ indicates stronger localization near the attractant maximum, whereas smaller values signify an inability to distinguish between high and low concentration regions.

\begin{figure}[hbt!]
\centering
\includegraphics[width=0.45\linewidth]{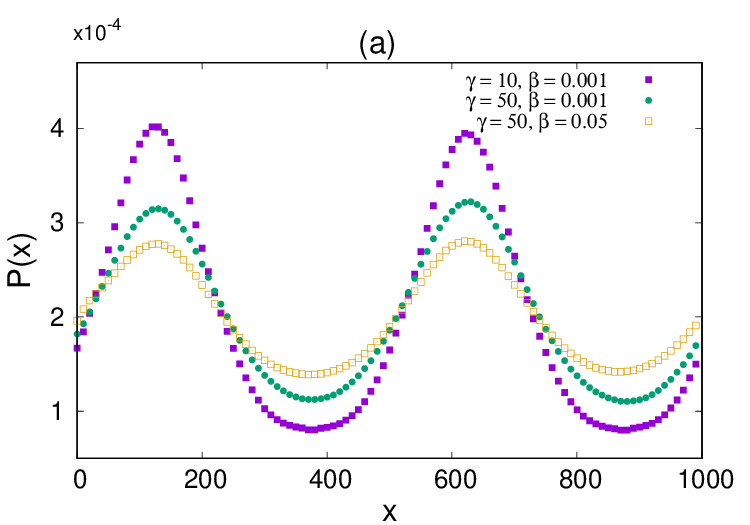}
\includegraphics[width=0.45\linewidth]{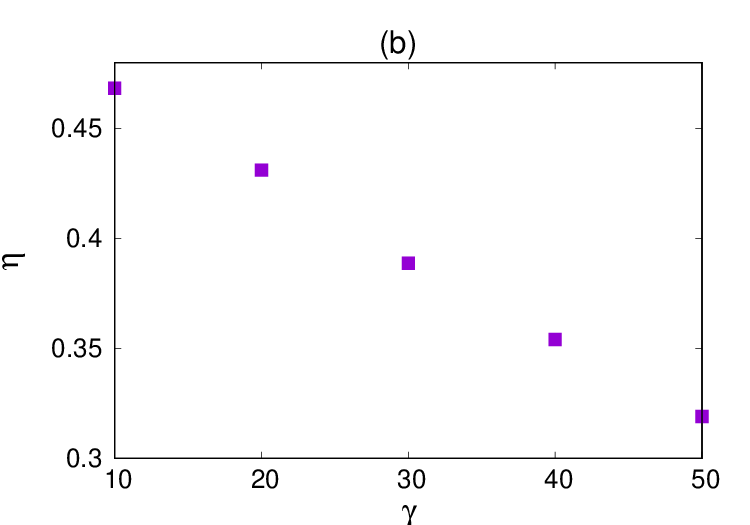}
\includegraphics[width=0.45\linewidth]{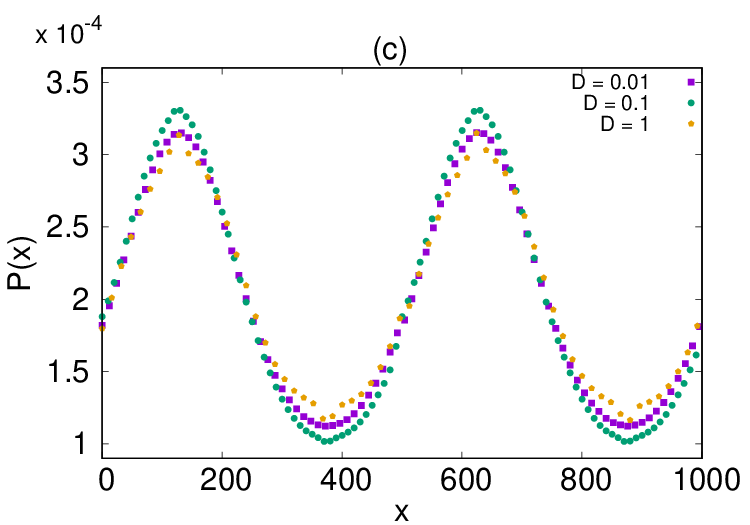}
\includegraphics[width=0.45\linewidth]{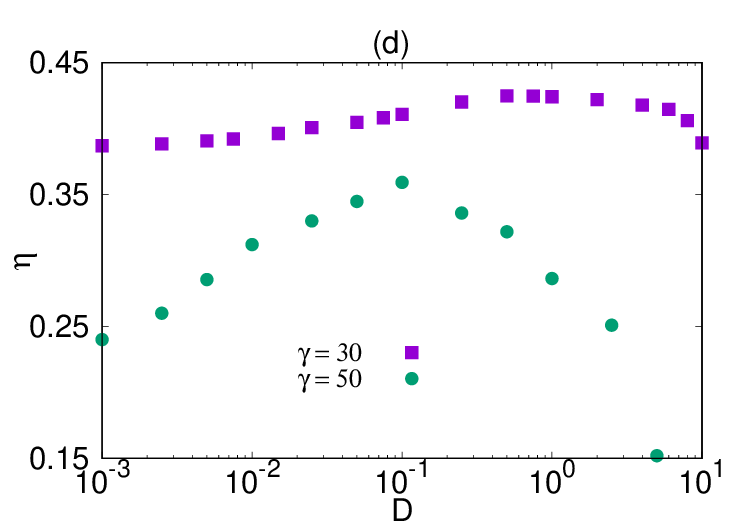}
\caption{Steady-state position distribution $P(x)$ of the RL agent in a sinusoidal chemical landscape, starting from a uniform initial distribution of the agent’s position. (a) $P(x)$ for different values of the chemical deposition rate $\gamma$ and evaporation rate $\beta$. Increasing $\gamma$ and $\beta$ progressively suppresses localization of the agent near the concentration maxima.  Here we have used $D = 0.01 s^{-1}$. (b) Contrast measure as a function of chemical deposition rate $\gamma$. Here, $D = 0.01 s^{-1}$ and $\beta = 0.001 s^{-1}$. (c) $P(x)$ for various values of $D$. Localization of the agent at the concentration hills first increases with D and subsequently decreases, exhibiting a nonmonotonic dependence. Here, $\gamma = 50 s^{-1}$ and $\beta = 0.001 s^{-1}$. (d) Variation of the contrast measure with $D$ for $\beta = 0.001 s^{-1}$. In all panels, the learning rate and exploration rate are fixed at $\alpha = 0.1$ and $\epsilon = 0.1$. The spatial coordinate $x$ is measured in micron ($\mu$m), $D$ and $\gamma$ are in $s^{-1}$.}
\label{FIG:sin_land_uniform}
\end{figure}

In Fig.~\ref{FIG:sin_land_uniform}, we present the steady-state position distribution $P(x)$ of the agent for various values of the deposition rate $\gamma$, evaporation rate $\beta$, and chemical hopping rate $D$, using a uniform initial condition for the agent’s position. The competition between the pre-existing attractant gradient in the medium and the gradient created by self-deposition controls the long time behavior of the agent. Fig. \ref{FIG:sin_land_uniform}(a) shows $P(x)$ for different choices of $\gamma$ and $\beta$, while other parameters are kept constant. For small $\gamma$ the initial attractant gradient guides the agent towards the nearest peak but as $\gamma$ becomes large, the self-generated chemical field dominates and tends to trap the agent near its initial position (which is equally likely to be anywhere in the system). This reduces the contrast in $P(x)$. In Fig. \ref{FIG:sin_land_uniform}(b) we plot the contrast $\eta$ as a function of $\gamma$ and verify this. Increasing $\beta$ on the other hand, causes the chemical to evaporate rapidly, which decreases the  gradient. This also causes the contrast to decrease, as seen in Fig. \ref{FIG:sin_land_uniform}(a). 

In Fig.~\ref{FIG:sin_land_uniform}(c) $P(x)$ is shown for different values of chemical hopping rate $D$. The contrast in $P(x)$ first increases with $D$ and then decreases. Fig. \ref{FIG:sin_land_uniform}(d) shows the contrast as a function of $D$ and we find a peak as in the previous subsection. The non-monotonic dependence of the contrast on D follows the same mechanism discussed in the previous subsection. We have also investigated (data not shown) the performance by varying the learning rate $\alpha$ and exploration rate $\epsilon$ and find that the contrast increases with $\alpha$ and decreases with $\epsilon$ monotonically, as discussed in the previous subsection.

\subsection{\textbf{ Multiple  attractant peaks of unequal heights}}

\begin{figure}[hbt!]
\centering
\includegraphics[width=0.45\linewidth]{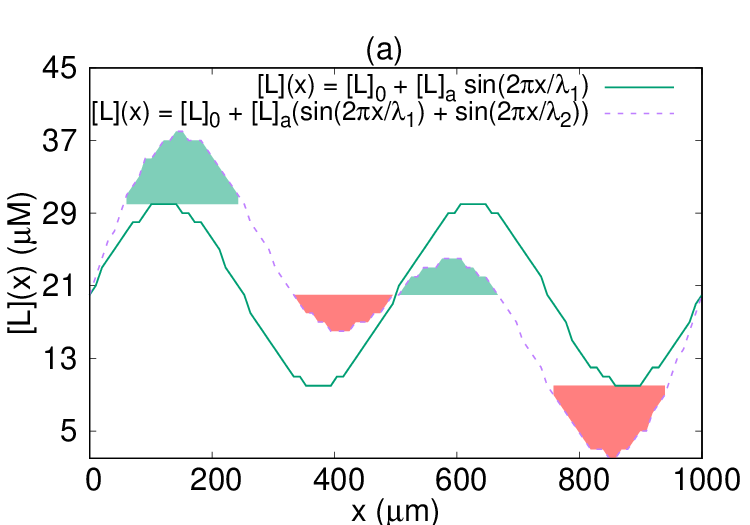}
\includegraphics[width=0.45\linewidth]{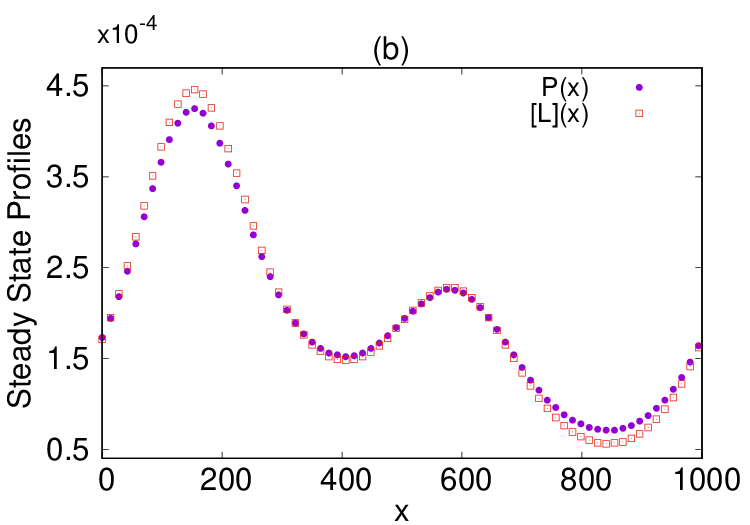}
\includegraphics[width=0.45\linewidth]{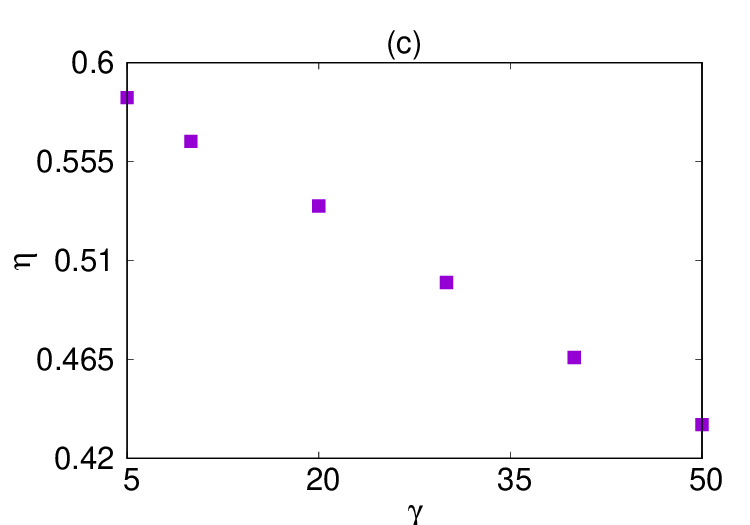}
\includegraphics[width=0.45\linewidth]{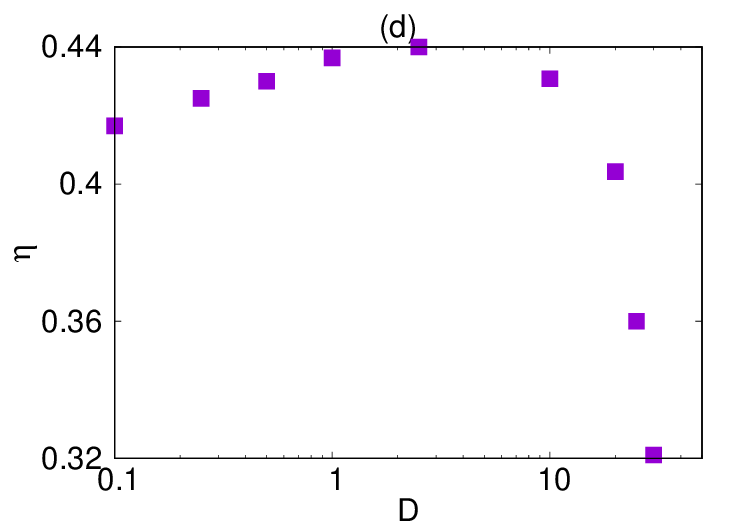}
\caption{ (a) Initial attractant concentration profile $[L](x,0)$, defined as the initial number of attractant molecules at position $x$. We use $[L]_0 + [L]_a \sin (2 \pi x / \lambda_1) $ to model an attractant profile with multiple peaks of equal height (solid line) and $[L]_0 + [L]_a \sin (2 \pi x / \lambda_1) + [L]_a \sin (2 \pi x / \lambda_2) $ for unequal heights (dashed line), with $[L]_0 = 20$, $[L]_a = 10 $, $\lambda_1 = 500 \mu m$, and $L=\lambda_2 = 1000 \mu m$. To ensure integer values at all $x$, we convert these functions to their nearest integer values. The blue-shaded regions indicate local attractant maxima, while the red-shaded regions indicate local attractant minima. We use periodic boundary conditions. (b) Steady-state profile $[L](x)$ of the attractant and the agent position distribution $P(x)$ . In this plot we have used $\gamma = 50 s^{-1}$, $\beta = 0.001 s^{-1}$, and $D = 0.01 s^{-1}$. (c) Contrast measure $\eta$ as a function of $\gamma$. Here $\beta = 0.001 s^{-1}$ and $D = 0.1 s^{-1}$. (d) Steady-state $\eta$ versus $D$, showing non-monotonic variation. Here, $\gamma = 50 s^{-1}$ and $\beta = 0.001 s^{-1}$. In all panels, the learning rate and exploration rate are fixed at $\alpha = 0.1$ and $\epsilon = 0.1$. The spatial coordinate $x$ is measured in micron ($\mu$m), $D$ and $\gamma$ are in $s^{-1}$.} \label{FIG:2SIN}
\end{figure}

To model an initial attractant profile with multiple peaks of unequal heights, we consider $[L](x,0)= [L]_0 + [L]_a \left[ \sin (2 \pi x / \lambda) +\sin ( \pi x / \lambda) \right ] $ with $\lambda=L/2$. This profile is shown in Fig. \ref{FIG:2SIN}(a). The RL agent starts from a random position drawn from uniform distribution. The long time position distribution $P(x)$ along with long time attractant profile $[L](x)$ are presented in Fig. \ref{FIG:2SIN}(b) which shows in the long time limit spatial variation of $P(x)$ matches with that of $[L](x)$ which is expected for a successful chemotaxis strategy. We investigate the chemotactic performance for different parameters. Just like we had shown in the previous subsection, with increasing deposition rate $\gamma$, the agent is less sensitive to the pre-existing gradient and responds more strongly to its own chemical trail, which is why the contrast $\eta$ decreases steadily with $\gamma$. The peak of $\eta$ with chemical hopping rate $D$ can also be explained by similar reasoning outlined in the previous subsection.

\begin{figure}[hbt!]
    \centering
    \includegraphics[width=0.95\linewidth]{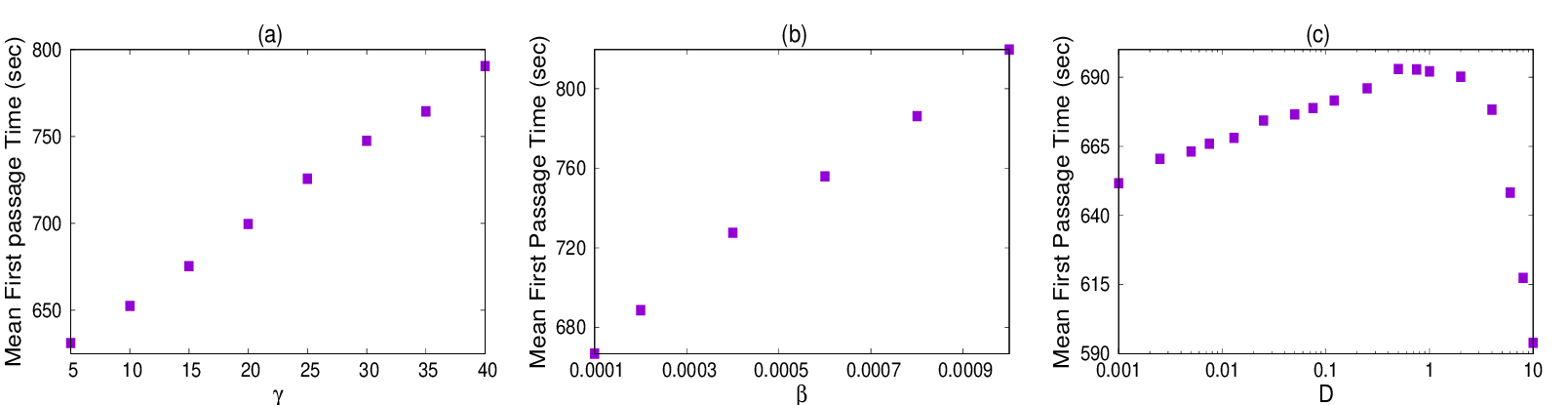}
   \caption{Mean first-passage time (MFPT) for an agent starting from the lower attractant peak and reaching the higher peak. (a) MFPT as a function of the chemical deposition rate $\gamma$ with fixed $\beta=0.001 s^{-1}$ and $D = 0.001 s^{-1}$. (b) MFPT as a function of the chemical evaporation rate $\beta$. Here,  $\gamma=10 s^{-1}$ and $D = 0.001 s^{-1}$. and  (c) MFPT as a function of $D$. Here, $\gamma=10 s^{-1}$ and $\beta = 0.001 s^{-1}$. In all panels, the learning rate and exploration rate are fixed at $\alpha = 0.1$ and $\epsilon = 0.1$. The MFPT is averaged over 2000 independent realizations. The parameters $\gamma$, $\beta$ and $D$ are measured in $s^{-1}$.} \label{FIG:2sin_MFPT_chemical}
\end{figure}

One important performance criterion in the case of multiple unequal peaks in the attractant environment is the search time. How long the agent needs to find the highest attractant peak is a question which becomes particularly important in presence of chemical degradation. In a harsh nutrient environment when the nutrient degrades or diffuses away fast, the chemotactic agent often needs to find the favorable region as soon as possible and make use of the nutrient while it lasts. In such situations, rather than the long time position distribution, the first passage time is a more suitable measure of performance. A good performance is indicated by short first passage time. We measure mean first passage time (MFPT) of the agent at the highest attractant peak starting from a smaller peak. More specifically, starting from the local peak, we measure how long does it take for the run-tumble trajectory to hit the global attractant peak for the first time. MFPT was calculated by averaging over many such trajectories. In Fig. \ref{FIG:2sin_MFPT_chemical}(a) we show MFPT increases as a function of $\gamma$ because with higher deposition rate the agent spends more time getting trapped near its initial position. MFPT also increases with degradation rate $\beta$ because in absence of any chemical cue the agent navigates through the system only via diffusion and MFPT takes a large value in that case (Fig. \ref{FIG:2sin_MFPT_chemical}(b)). Fig. \ref{FIG:2sin_MFPT_chemical}(c) shows a peak in MFPT as a function of $D$. Going from a local maximum to the global maximum of attractant profile requires the agent to go downhill and then uphill along the pre-existing gradient. A large portion of the time is spent escaping the local maximum while climbing uphill to reach the global peak is relatively quicker. For small $D$ the trapping of the agent at the local maximum is caused only by the pre-existing gradient. But as $D$ increases slightly, the agent experiences the gradient of self-secreted attractant as well. The trapping is stronger now and this increases MFPT. For much larger $D$ gradient is weaker and it is easier for the agent to escape the trap. This decreases MFPT. Even if the agent starts from some other position, like the minimum of the attractant profile, it can move in either direction to find higher attractant level. But one direction takes it towards the local peak, and the other towards the global peak. Even in this case, those trajectories which take the agent closer to the local peak, have to first escape the influence of the local peak and then find the global peak. We have verified that MFPT shows similar trend in this case (data not shown).

\begin{figure}[hbt!]
    \centering
    \includegraphics[width=0.4\linewidth]{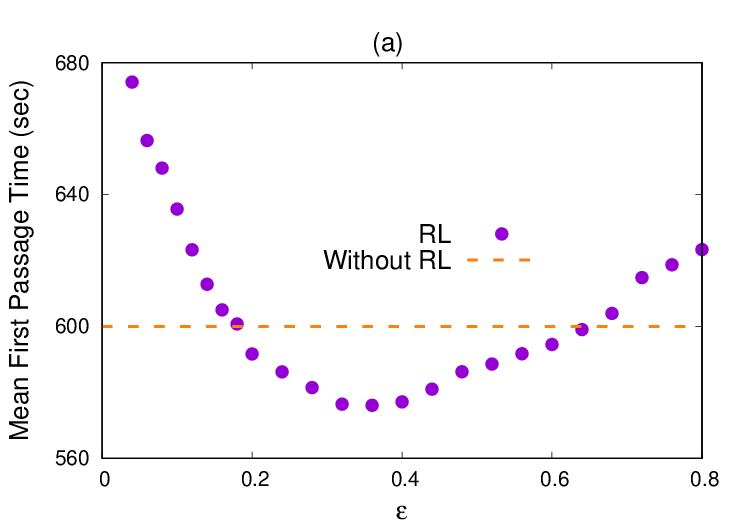}
    \includegraphics[width=0.4\linewidth]{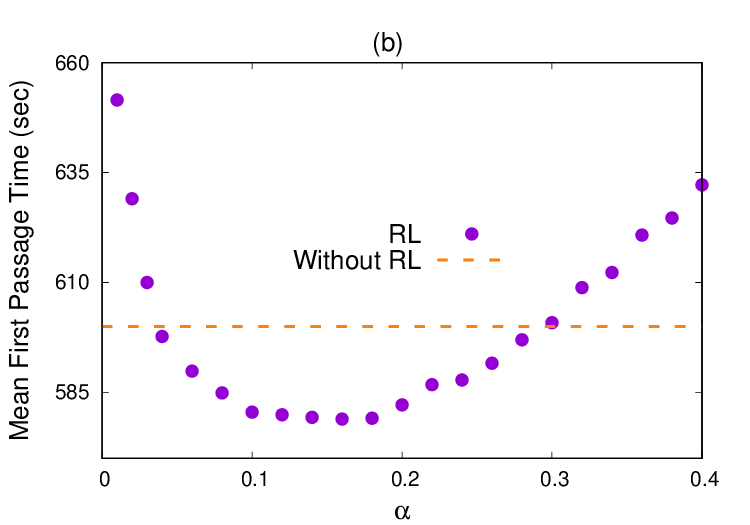}
   \caption{Mean first-passage time (MFPT) for an agent starting from the lower attractant peak and reaching the higher peak. (a) MFPT as a function of exploration rate $\epsilon$ for $\alpha = 0.1$.  (b) MFPT as a function of learning rate $\alpha$ with exploration rate fixed at $\epsilon = 0.3$. The orange lines in panels (a) and (b) indicate the MFPT obtained without RL {\sl i.e.} using   a standard model for chemotaxis where a non-Markovian run-and-tumble motion is considered, as explained in Sec. \ref{sec:model}D. Here we have used same $[L](x,0)$ and same $[L](x,t)$ time-evolution as those used in the RL case. The MFPT is averaged over 2000 independent realizations. In both the panels we have used $\gamma = 5 s^{-1}$, $\beta = 0.005 s^{-1}$ and $D = 0.001 s^{-1}$.} \label{FIG:MFPT}
\end{figure}

We also investigate how exploration and exploitation influence MFPT. For small exploration rate $\epsilon$ the RL agent attempts to exploit its recent learning which prevents it from exploring its environment. This keeps the agent confined near the local peak for long time, leading to large MFPT. In contrast, in the limit $\epsilon \to 1$, the affinity for attractant peaks is strongly reduced and the agent’s motion becomes nearly diffusive, again resulting in slow searches with large MFPT. Between these two extremes, the MFPT exhibits a pronounced minimum [Fig.~\ref{FIG:MFPT}(a)], signaling the existence of an optimal level of exploration. At fixed exploration strength, we also examine the variation of MFPT with the learning rate $\alpha$ [Fig.~\ref{FIG:MFPT}(b)], and find that an intermediate $\alpha$ likewise minimizes the search time, highlighting the presence of an optimal learning strategy.

\section{Reinforcement Learning Yields Better Performance}
\label{sec:nmrw}

To estimate how effective the RL strategy is, we compare the performance of the RL agent with a standard model for chemotaxis where a run-and-tumble motion is considered with tumbling rate depending on recent memory \cite{Gennes,kafri,PhysRevE.91.042714}, as described in Eq. \ref{eq:memory}. The dynamics of the attractant field remains the same with deposition, degradation and diffusion, as described in Sec. \ref{sec:model}D. 

\begin{figure}
\includegraphics[scale=0.46]{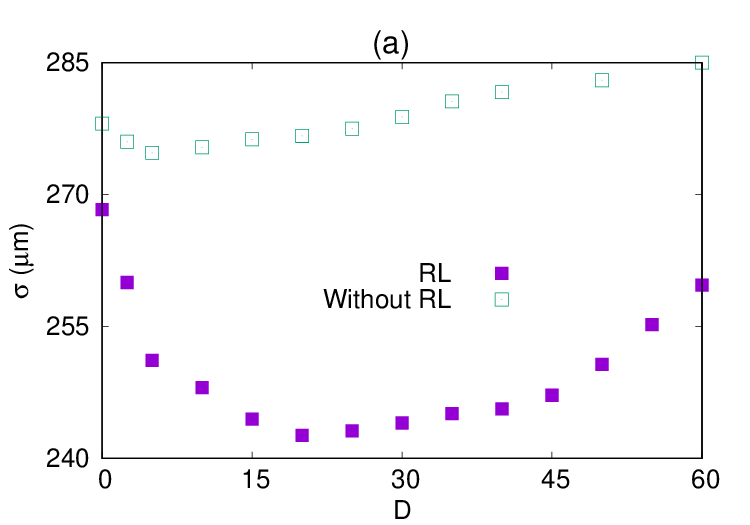}
\includegraphics[scale=0.46]{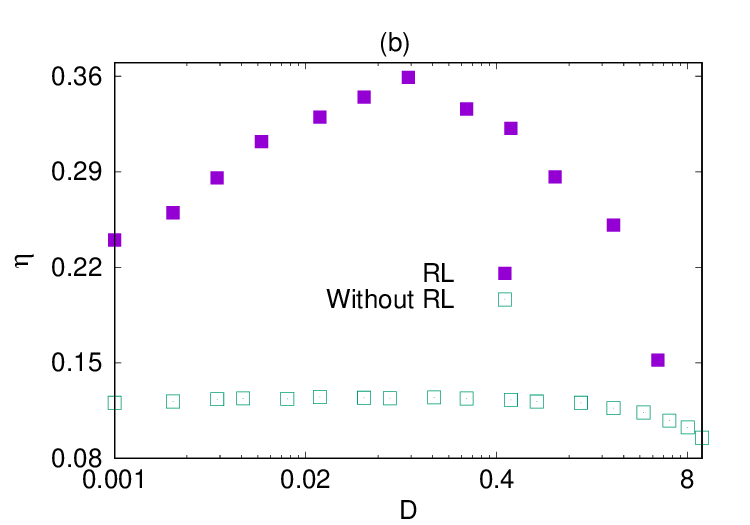}
\includegraphics[scale=0.46]{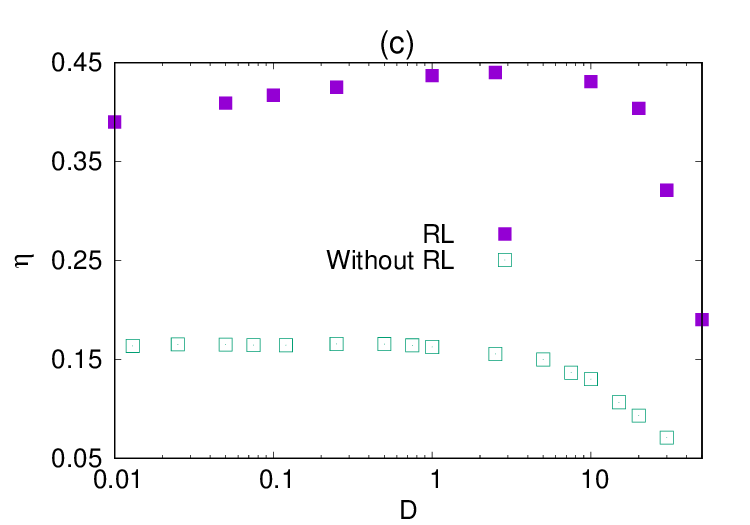}
\caption{Performance comparison between the RL and non-RL models for three different chemical environments. (a) For a flat initial concentration profile, the width $\sigma$ of the steady-state distribution $P(x)$ is plotted versus $D$. The RL model (red) yields smaller $\sigma$ than the non-RL case (orange), indicating stronger localization. Here, $\gamma = 50 s^{-1}$ and $\beta = 0.005 s^{-1}$. (b) Contrast $\eta$ versus $D$ for a sinusoidal environment, with $\gamma = 50 s^{-1}$ and $\beta = 0.001 s^{-1}$. (c) Contrast measure $\eta$ as a function of $D$ for a chemical environment with multiple peaks of unequal heights. We take $\gamma = 50 s^{-1}$ and $\beta = 0.001 s^{-1}$. In panels (b) and (c), $\eta$ is larger in the RL model (red) than in the non-RL case (orange), indicating enhanced localization near chemical concentration hills.  For the RL case the exploration and learning parameters are fixed at $\epsilon = 0.1$ and $\alpha = 0.1$. Here $D$ is measured in units of $s^{-1}$.
} \label{fig:eff}
\end{figure}

\begin{figure}[hbt!]
    \centering
    \includegraphics[width=0.45\linewidth]{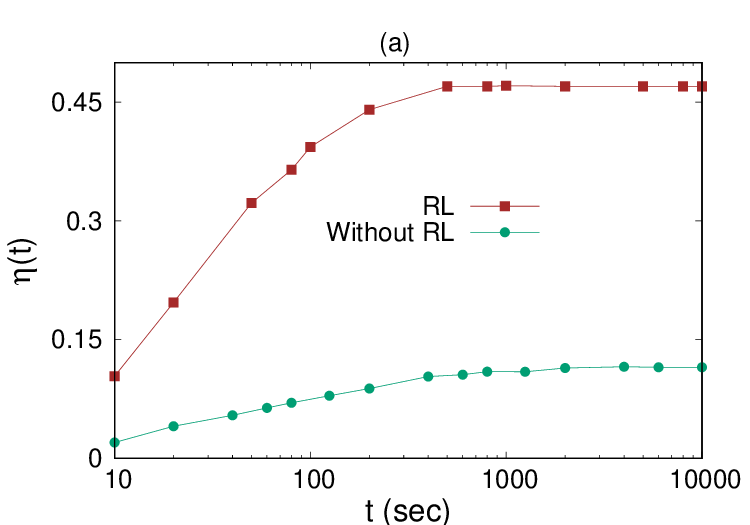}
    \includegraphics[width=0.45\linewidth]{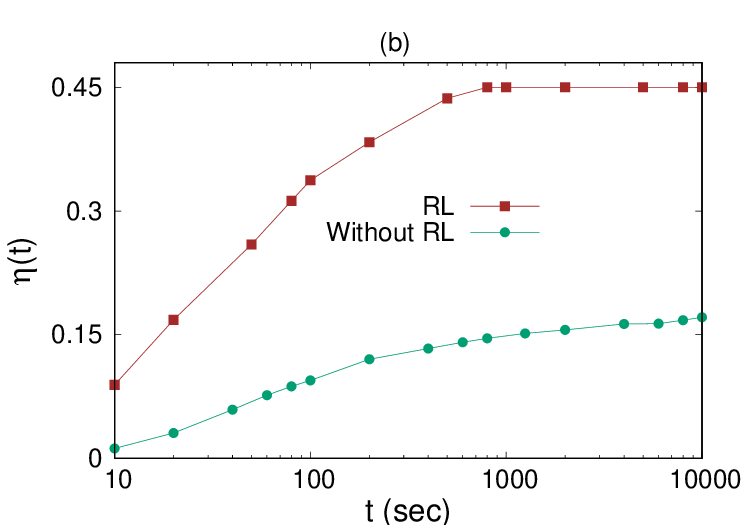}
  \caption{Temporal variation of the contrast measure $\eta$ for models with RL agent (brown) and without RL (green) for (a) a sinusoidal initial chemical profile and (b) an initial chemical profile consisting of multiple peaks with different heights. For the RL case the exploration and learning parameters are fixed at $\epsilon = 0.5$ and $\alpha = 0.5$ such that the mean run-duration of the RL-agent becomes same with that of non-Markov random walker. In both the panels we have used $\gamma = 10 s^{-1}$, $\beta = 0.001 s^{-1}$ and $D = 0.1 s^{-1}$. Here time $t$ is measured in seconds ($s$). } \label{FIG:eta_time}
\end{figure}

For an attractant profile that was initially flat, a measure of performance is standard deviation $\sigma$ of the steady state position. An optimum diffusivity of the attractant chemical  can bring $\sigma$ down to a minimum value. In Fig. \ref{fig:eff}(a) we compare the $\sigma$ vs $D$ variation for our RL agent and the run-tumble model without RL. We find a significantly better performance (lower $\sigma$) when RL is employed. For the case when the initial attractant profile has sinusoidal variation, contrast $\eta$ is used as a marker for chemotactic performance which shows a highest value at optimal diffusivity. In Fig. \ref{fig:eff}(b) we compare $\eta$ from two models across a range of $D$ and find that higher $\eta$ can be achieved using RL. Same conclusions can be reached for an attractant profile with superposition of two sine waves (see Fig. \ref{fig:eff}(c)). Not only RL helps reach a higher contrast in steady state, it also reaches that value faster. In Fig. \ref{FIG:eta_time} we plot the contrast as a function of time for the agent using RL and without RL. For a uniform initial position, the initial contrast is zero. We find $\eta$ increases from this value and saturates to its steady state value for both models. But the rise is much faster for the model with RL. Another performance criterion used in this work was search time. Even in this case by choosing the exploitation rate $\epsilon$ and learning rate $\alpha$ it was possible to bring MFPT down to an optimal value which was lower than MFPT obtained for an agent without RL (see Fig. \ref{FIG:MFPT}).

Therefore, we conclude that using RL is generally beneficial for the agent. It helps improve its long time performance, makes it possible for the agent to find the high attractant zone quicker. RL also brings down the relaxation time of the system, which can have a computational advantage too.

\section{Conclusions}  \label{sec:con}

In this work, we have studied a run-tumble motion with autochemotaxis using a reinforcement learning framework. While the RL agent senses and responds to its chemical environment, it also modifies the environment by continuously secreting the same chemical. The chemical diffuses and degrades in time, and the agent senses the resulting concentration field to guide its motion. A successful RL strategy enables the agent to be present near the attractant peaks with large probability. We show that employing RL indeed yields better chemotactic performance compared to usual run-tumble chemotaxis without RL. In the long time limit an RL agent shows higher occupancy near the attractant peaks. Moreover, tracking this occupancy with time reveals that not only the occupancy reaches a higher value for an RL agent, it also reaches this value faster. Thus RL can be employed as an efficient and successful navigation strategy in autotaxis.

The behavior of the RL agent is a result of interplay between the self-generated chemical field and the  pre-existing chemical field in the medium. The initial chemical concentration can be flat, or can have one or more peaks. In case of multiple peaks they can be of equal and unequal heights. Based on this, we have considered three generic initial chemical concentration: a flat profile, a sinusoidal profile and superposition of two sine waves. While a sinusoidal profile has multiple peaks of equal height, the superposition of two sine waves gives rise to multiple peaks of unequal height for which the agent may experience both local and global maxima. However, our conclusions are not sensitive to the specific functional form chosen and will remain valid for any other choice of function. For these different choices of initial chemical field, we quantitatively characterize the chemotactic performance by varying different parameters like rates of secretion, degradation and diffusion of the chemical. Apart from the long time performance, RL also enables the agent to search for the attractant peak efficiently. By striking a balance between the learning and exploration, an RL agent can find the attractant peak in shortest possible time. This strategy can be directly implemented in robotic RL systems to carry out efficient chemotactic search \cite{7hcf-p1yk}.

Throughout this work, we have restricted our analysis to one spatial dimension. In addition to simplicity, this choice is motivated by experiments on {\sl E. coli} performed in narrow microfluidic channels, where the channel width is comparable to or smaller than the typical run length and the motion is effectively one-dimensional \cite{PhysRevLett.118.098101,10.1073/pnas.1108608108,PhysRevLett.108.128101}. A natural extension of our framework is to study higher-dimensional systems, where a tumbling event does not reverse the run direction, but randomizes it. It will be of interest to investigate the effectiveness of RL algorithm in such a setting. In particular, earlier studies on passive Brownian particles interacting with self-generated chemical fields have shown that self-trapping via autochemotaxis occurs in one and two dimensions, but not in three dimensions \cite{Tsori_2004,PhysRevE.80.031122}. It is an intriguing question whether an active run-tumble motion and a suitably formulated RL strategy can change these conclusions.

It is also worth noting that we deliberately employ a minimal state-action space consisting of four state-action pairs: two states, high and low attractant concentration, and two actions, persist and reverse. This minimal representation is sufficient to capture the essential decision-making mechanism of an E. coli cell while keeping the model computationally tractable. Similar minimal state-action representations have also been employed in previous RL studies \cite{PhysRevE.104.054614,Landin,Wang_2021,D5SM01040C}.

Previous studies also report rich collective behavior in systems involving multiple interacting autotactic cells \cite{PhysRevLett.131.088201, Amar2016, PhysRevE.98.052411, Livne, Mukherjee, Ma2016}. Our model can be extended to capture some of these phenomena. Finally, few generalizations of our simple RL algorithm can be attempted. Instead of working with fixed learning parameters, one might allow these parameters to change dynamically, such that the agent can continuously adjust its learning strategy as the system evolves in time. A sophisticated RL algorithm like this may prove beneficial for the agent. Possible experimental realizations include active colloids or run-and-tumble microrobots equipped with learning-based control, which could provide a direct test of the predictions of our model.

\section{Acknowledgements}
RP acknowledges the research fellowship [Grant No. 09/0575(13013)/2022-EMR-I] from the Council of Scientific and Industrial Research (CSIR), India. SC acknowledges support from Anusandhan National Research Foundation (ANRF), India (Grant No: CRG/2023/000159).


\begin{thebibliography}{10}

\bibitem{eisenbachbook}
Michael Eisenbach.
\newblock {\em Chemotaxis}.
\newblock Published by Imperial College Press and Distributed by World
  Scientific Publishing Co., 2004.

\bibitem{berg2008coli}
Howard~C Berg.
\newblock {\em \textit{E. coli} in Motion}.
\newblock Springer Science \& Business Media, 2008.

\bibitem{Moore}
Jeremy~P. Moore and Thierry Emonet.
\newblock Physics of bacterial chemotaxis.
\newblock {\em Current Biology}, 34(20):R972--R977, 2024.

\bibitem{Young}
L.~Y. Young and Ralph Mitchell.
\newblock Negative chemotaxis of marine bacteria to toxic chemicals.
\newblock {\em Applied Microbiology}, 25(6):972--975, 1973.

\bibitem{Vargas}
Pablo Vargas, Lucie Barbier, Pablo~José Sáez, and Matthieu Piel.
\newblock Mechanisms for fast cell migration in complex environments.
\newblock {\em Current Opinion in Cell Biology}, 48:72--78, 2017.
\newblock Cell Dynamics.

\bibitem{Friedrich}
Benjamin~M. Friedrich and Frank Jülicher.
\newblock Chemotaxis of sperm cells.
\newblock {\em Proceedings of the National Academy of Sciences},
  104(33):13256--13261, 2007.

\bibitem{Hamdoun}
Héctor~Vicente Ramírez-Gómez, Idán Tuval, Adán Guerrero, and Alberto
  Darszon.
\newblock Chapter 21 - analysis of sperm chemotaxis.
\newblock In Amro Hamdoun and Kathy~R. Foltz, editors, {\em Echinoderms, Part
  B}, volume 151 of {\em Methods in Cell Biology}, pages 473--486. Academic
  Press, 2019.

\bibitem{Kaupp}
U.~Benjamin Kaupp, Nachiket~D. Kashikar, and Ingo Weyand.
\newblock Mechanisms of sperm chemotaxis.
\newblock {\em Annual Review of Physiology}, 70(Volume 70, 2008):93--117, 2008.

\bibitem{Eisenbach2025}
Michael Eisenbach.
\newblock Sperm navigation in humans: a concerted action of multiple means.
\newblock {\em Communications Biology}, 8(1):923, 2025.

\bibitem{Lange}
Steffen Lange and Benjamin~M. Friedrich.
\newblock Sperm chemotaxis in marine species is optimal at physiological flow
  rates according theory of filament surfing.
\newblock {\em PLOS Computational Biology}, 17(4):1--16, 04 2021.

\bibitem{PhysRevLett.95.128103}
R.~Grima.
\newblock Strong-coupling dynamics of a multicellular chemotactic system.
\newblock {\em Phys. Rev. Lett.}, 95:128103, Sep 2005.

\bibitem{PhysRevE.80.031122}
Ankush Sengupta, Sven van Teeffelen, and Hartmut L\"owen.
\newblock Dynamics of a microorganism moving by chemotaxis in its own
  secretion.
\newblock {\em Phys. Rev. E}, 80:031122, Sep 2009.

\bibitem{Mokhtari}
Zahra Mokhtari, Robert I~A Patterson, and Felix Höfling.
\newblock Spontaneous trail formation in populations of auto-chemotactic
  walkers.
\newblock {\em New Journal of Physics}, 24(1):013012, jan 2022.

\bibitem{Konijn}
T~M Konijn, J~G Van~De Meene, J~T Bonner, and D~S Barkley.
\newblock The acrasin activity of adenosine-3',5'-cyclic phosphate.
\newblock {\em Proceedings of the National Academy of Sciences},
  58(3):1152--1154, 1967.

\bibitem{Gregor}
Thomas Gregor, Koichi Fujimoto, Noritaka Masaki, and Satoshi Sawai.
\newblock The onset of collective behavior in social amoebae.
\newblock {\em Science}, 328(5981):1021--1025, 2010.

\bibitem{Afonso}
Philippe V. Afonso, Mirkka Janka-Junttila, Young Jong Lee, Colin P. McCann,
  Charlotte M. Oliver, Khaled A. Aamer, Wolfgang Losert, Marcus T. Cicerone,
  and Carole A. Parent.
\newblock Ltb4 is a signal-relay molecule during neutrophil chemotaxis.
\newblock {\em Developmental Cell}, 22(5):1079--1091, 2012.

\bibitem{Budrene}
Elena~O. Budrene and Howard~C. Berg.
\newblock Complex patterns formed by motile cells of escherichia coli.
\newblock {\em Nature}, 349(6310):630--633, 1991.

\bibitem{Tsori_2004}
Y.~Tsori and P.-G. de~Gennes.
\newblock Self-trapping of a single bacterium in its own chemoattractant.
\newblock {\em Europhysics Letters}, 66(4):599, may 2004.

\bibitem{PhysRevLett.131.088201}
Jasper van~der Kolk, Florian Ra\ss{}hofer, Richard Swiderski, Astik Haldar,
  Abhik Basu, and Erwin Frey.
\newblock Anomalous collective dynamics of autochemotactic populations.
\newblock {\em Phys. Rev. Lett.}, 131:088201, Aug 2023.

\bibitem{Amar2016}
M.~Ben Amar.
\newblock Collective chemotaxis and segregation of active bacterial colonies.
\newblock {\em Scientific Reports}, 6(1):21269, 2016.

\bibitem{PhysRevE.98.052411}
Enkeleida Lushi, Raymond~E. Goldstein, and Michael~J. Shelley.
\newblock Nonlinear concentration patterns and bands in autochemotactic
  suspensions.
\newblock {\em Phys. Rev. E}, 98:052411, Nov 2018.

\bibitem{Livne}
Nir Livne, Ady Vaknin, and Oded Agam.
\newblock Pattern formation in $e$. coli through negative chemotaxis:
  Instability, condensation, and merging.
\newblock {\em Phys. Rev. Res.}, 7:023095, Apr 2025.

\bibitem{Mukherjee}
Mrinmoy Mukherjee and Pushpita Ghosh.
\newblock Growth-mediated autochemotactic pattern formation in self-propelling
  bacteria.
\newblock {\em Phys. Rev. E}, 97:012413, Jan 2018.

\bibitem{Ma2016}
Manjun Ma, Meiyan Gao, Changqing Tong, and Yazhou Han.
\newblock Chemotaxis-driven pattern formation for a
  reaction--diffusion--chemotaxis model with volume-filling effect.
\newblock {\em Computers \& Mathematics with Applications}, 72(5):1320--1340,
  2016.

\bibitem{sutton2018reinforcement}
Richard~S Sutton and Andrew~G Barto.
\newblock {\em Reinforcement learning: An introduction}.
\newblock MIT press, 2018.

\bibitem{panait2005cooperative}
Liviu Panait and Sean Luke.
\newblock Cooperative multi-agent learning: The state of the art.
\newblock {\em Autonomous agents and multi-agent systems}, 11:387--434, 2005.

\bibitem{busoniu2008comprehensive}
Lucian Busoniu, Robert Babuska, and Bart De~Schutter.
\newblock A comprehensive survey of multiagent reinforcement learning.
\newblock {\em IEEE Transactions on Systems, Man, and Cybernetics, Part C
  (Applications and Reviews)}, 38(2):156--172, 2008.

\bibitem{kubat2017introduction}
Miroslav Kubat.
\newblock {\em An introduction to machine learning}.
\newblock Springer, 2017.

\bibitem{watkins1992q}
Christopher~JCH Watkins and Peter Dayan.
\newblock Q-learning.
\newblock {\em Machine learning}, 8:279--292, 1992.

\bibitem{NIPS2005_9752d873}
Martin Zinkevich, Amy Greenwald, and Michael Littman.
\newblock Cyclic equilibria in markov games.
\newblock In Y.~Weiss, B.~Sch\"{o}lkopf, and J.~Platt, editors, {\em Advances
  in Neural Information Processing Systems}, volume~18. MIT Press, 2005.

\bibitem{kanai}
Ildefons~Magrans de~Abril and Ryota Kanai.
\newblock Curiosity-driven reinforcement learning with homeostatic regulation.
\newblock In {\em 2018 International Joint Conference on Neural Networks,
  {IJCNN} 2018, Rio de Janeiro, Brazil, July 8-13, 2018}, pages 1--6. {IEEE},
  2018.

\bibitem{Mohan}
Prajval Mohan, Lakshya Sharma, and Pranav Narayan.
\newblock Optimal path finding using iterative sarsa.
\newblock In {\em 2021 5th International Conference on Intelligent Computing
  and Control Systems (ICICCS)}, pages 811--817, 2021.

\bibitem{schulman2017}
John Schulman, Filip Wolski, Prafulla Dhariwal, Alec Radford, and Oleg Klimov.
\newblock Proximal policy optimization algorithms, 2017.

\bibitem{sampat2022ordering}
Pranay~Bimal Sampat, Ananya Verma, Riya Gupta, and Shradha Mishra.
\newblock Ordering through learning in two-dimensional ising spins.
\newblock {\em Physical Review E}, 106(5):054149, 2022.

\bibitem{Kumar_2025}
Anish Kumar, Pawan~Kumar Mishra, Riya Singh, Shradha Mishra, and Debaprasad
  Giri.
\newblock Adaptive dynamics of ising spins in one dimension leveraging
  reinforcement learning.
\newblock {\em Physica Scripta}, 100(5):056009, apr 2025.

\bibitem{Neftci2019}
Emre~O. Neftci and Bruno~B. Averbeck.
\newblock Reinforcement learning in artificial and biological systems.
\newblock {\em Nature Machine Intelligence}, 1(3):133--143, 2019.

\bibitem{Treloar}
Neythen~J. Treloar, Nathan Braniff, Brian Ingalls, and Chris~P. Barnes.
\newblock Deep reinforcement learning for optimal experimental design in
  biology.
\newblock {\em PLOS Computational Biology}, 18(11):1--24, 11 2022.

\bibitem{Yongacoglu}
Bora Yongacoglu, Gürdal Arslan, and Serdar Yüksel.
\newblock Reinforcement learning for decentralized stochastic control.
\newblock In {\em 2019 IEEE 58th Conference on Decision and Control (CDC)},
  pages 5556--5561, 2019.

\bibitem{Zhu}
X.~Zhu, Y.~Wang, and Z.~Wu.
\newblock Reinforcement learning for optimal control of stochastic nonlinear
  systems.
\newblock {\em AIChE Journal}, 71, 2025.

\bibitem{MunosBourgine1997}
R{\'e}mi Munos and Paul Bourgine.
\newblock Reinforcement learning for continuous stochastic control problems.
\newblock In Michael~I. Jordan, Michael~J. Kearns, and Sara~A. Solla, editors,
  {\em Advances in Neural Information Processing Systems}, volume~10. MIT
  Press, 1997.

\bibitem{TEJEDOR2020101836}
Miguel Tejedor, Ashenafi~Zebene Woldaregay, and Fred Godtliebsen.
\newblock Reinforcement learning application in diabetes blood glucose control:
  A systematic review.
\newblock {\em Artificial Intelligence in Medicine}, 104:101836, 2020.

\bibitem{PhysRevX.9.031012}
Thomas~J. Rademaker, Emmanuel Bengio, and Paul Fran\ifmmode~\mbox{\c{c}}\else
  \c{c}\fi{}ois.
\newblock Attack and defense in cellular decision-making: Lessons from machine
  learning.
\newblock {\em Phys. Rev. X}, 9:031012, Jul 2019.

\bibitem{durve2020learning}
Mihir Durve, Fernando Peruani, and Antonio Celani.
\newblock Learning to flock through reinforcement.
\newblock {\em Physical Review E}, 102(1):012601, 2020.

\bibitem{Hou2019}
Hanxu Hou, Tian Gan, Yaodong Yang, Xianglei Zhu, Sen Liu, Weiming Guo, and
  Jianye Hao.
\newblock Using deep reinforcement learning to speed up collective cell
  migration.
\newblock {\em BMC Bioinformatics}, 20(18):571, 2019.

\bibitem{PhysRevE.111.014106}
Ramesh Pramanik, Shradha Mishra, and Sakuntala Chatterjee.
\newblock Run-and-tumble chemotaxis using reinforcement learning.
\newblock {\em Phys. Rev. E}, 111:014106, Jan 2025.

\bibitem{D5SM01040C}
Nicholas Tovazzi, Gorka Muñoz-Gil, and Michele Caraglio.
\newblock Run-and-tumble particles learning chemotaxis.
\newblock {\em Soft Matter}, 21:8886--8896, 2025.

\bibitem{Berg1972}
Howard~C. Berg and Douglas~A. Brown.
\newblock Chemotaxis in escherichia coli analysed by three-dimensional
  tracking.
\newblock {\em Nature}, 239(5374):500--504, 1972.

\bibitem{PhysRevE.99.012402}
Subrata Dev and Sakuntala Chatterjee.
\newblock Run-and-tumble motion with steplike responses to a stochastic input.
\newblock {\em Phys. Rev. E}, 99:012402, Jan 2019.

\bibitem{Gennes}
P.-G. de~Gennes.
\newblock Chemotaxis: the role of internal delays.
\newblock {\em European Biophysics Journal}, 33(8):691--693, 2004.

\bibitem{kafri}
Sakuntala Chatterjee, Rava~Azeredo da~Silveira, and Yariv Kafri.
\newblock Chemotaxis when bacteria remember: Drift versus diffusion.
\newblock {\em PLOS Computational Biology}, 7(12):1--8, 12 2011.

\bibitem{ladimirov}
Nikita Vladimirov, Dirk Lebiedz, and Victor Sourjik.
\newblock Predicted auxiliary navigation mechanism of peritrichously
  flagellated chemotactic bacteria.
\newblock {\em PLOS Computational Biology}, 6:1--8, 03 2010.

\bibitem{PhysRevE.97.032420}
Subrata Dev and Sakuntala Chatterjee.
\newblock Optimal methylation noise for best chemotactic performance of e.
  coli.
\newblock {\em Phys. Rev. E}, 97:032420, Mar 2018.

\bibitem{PhysRevE.105.014411}
Shobhan~Dev Mandal and Sakuntala Chatterjee.
\newblock Effect of receptor cooperativity on methylation dynamics in bacterial
  chemotaxis with weak and strong gradient.
\newblock {\em Phys. Rev. E}, 105:014411, Jan 2022.

\bibitem{Mandal_2025}
Shobhan~Dev Mandal and Sakuntala Chatterjee.
\newblock Bacterial chemotaxis in a traveling wave attractant environment.
\newblock {\em Journal of Statistical Mechanics: Theory and Experiment},
  2025(6):063501, jun 2025.

\bibitem{Chatterjee_2022}
Sakuntala Chatterjee.
\newblock Short time extremal response to step stimulus for a single cell e.
  coli.
\newblock {\em Journal of Statistical Mechanics: Theory and Experiment},
  2022(12):123503, dec 2022.

\bibitem{Pramanik2025}
Ramesh Pramanik, Ramu~K. Yadav, and Sakuntala Chatterjee.
\newblock Dynamics of chemoreceptor activity with time-periodic attractant
  field.
\newblock {\em The European Physical Journal E}, 48(10):60, 2025.

\bibitem{Mandal2022}
Shobhan~Dev Mandal and Sakuntala Chatterjee.
\newblock Effect of switching time scale of receptor activity on chemotactic
  performance of {Escherichia coli}.
\newblock {\em Indian Journal of Physics}, 96(9):2619--2627, 2022.

\bibitem{PhysRevE.103.L030401}
Shobhan~Dev Mandal and Sakuntala Chatterjee.
\newblock Effect of receptor clustering on chemotactic performance of e. coli:
  Sensing versus adaptation.
\newblock {\em Phys. Rev. E}, 103:L030401, Mar 2021.

\bibitem{Mingshan}
Mingshan Li and Gerald~L. Hazelbauer.
\newblock Cellular stoichiometry of the components of the chemotaxis signaling
  complex.
\newblock {\em Journal of Bacteriology}, 186(12):3687--3694, 2004.

\bibitem{Tedesco}
H~C Berg and P~M Tedesco.
\newblock Transient response to chemotactic stimuli in escherichia coli.
\newblock {\em Proceedings of the National Academy of Sciences},
  72(8):3235--3239, 1975.

\bibitem{PhysRevLett.100.238101}
Yariv Kafri and Rava~Azeredo da~Silveira.
\newblock Steady-state chemotaxis in escherichia coli.
\newblock {\em Phys. Rev. Lett.}, 100:238101, Jun 2008.

\bibitem{PhysRevE.91.042714}
Subrata Dev and Sakuntala Chatterjee.
\newblock Optimal search in e. coli chemotaxis.
\newblock {\em Phys. Rev. E}, 91:042714, Apr 2015.

\bibitem{Segall}
J~E Segall, S~M Block, and H~C Berg.
\newblock Temporal comparisons in bacterial chemotaxis.
\newblock {\em Proceedings of the National Academy of Sciences},
  83(23):8987--8991, 1986.

\bibitem{Block1}
S~M Block, J~E Segall, and H~C Berg.
\newblock Adaptation kinetics in bacterial chemotaxis.
\newblock {\em Journal of Bacteriology}, 154(1):312--323, 1983.

\bibitem{Brown}
Douglas~A. Brown and Howard~C. Berg.
\newblock Temporal stimulation of chemotaxis in \textit{Escherichia coli}.
\newblock {\em Proceedings of the National Academy of Sciences},
  71(4):1388--1392, 1974.

\bibitem{Macnab}
Robert~M. Macnab and D.~E. Koshland.
\newblock The gradient-sensing mechanism in bacterial chemotaxis.
\newblock {\em Proceedings of the National Academy of Sciences},
  69(9):2509--2512, 1972.

\bibitem{Putzke2023}
Mischa Putzke and Holger Stark.
\newblock Optimal navigation of a smart active particle: directional and
  distance sensing.
\newblock {\em The European Physical Journal E}, 46(6):48, 2023.

\bibitem{Monderkamp_2022}
Paul~A Monderkamp, Fabian~Jan Schwarzendahl, Michael~A Klatt, and Hartmut
  Löwen.
\newblock Active particles using reinforcement learning to navigate in complex
  motility landscapes.
\newblock {\em Machine Learning: Science and Technology}, 3(4):045024, dec
  2022.

\bibitem{Man2025}
Hai Man, Chaobo Wang, Jia-Rui Li, Yuping Tian, and Shu-Gang Chen.
\newblock Autonomous discovery of the ising model’s critical parameters with
  reinforcement learning.
\newblock {\em Journal of Statistical Mechanics: Theory and Experiment},
  2025(12):123410, dec 2025.

\bibitem{Zai2026}
Anja~T. Zai, Corinna Lorenz, Shakana Srikantharajah, Nicolas Giret, and Richard
  H.~R. Hahnloser.
\newblock Estimating the motor exploration in reinforcement learning.
\newblock {\em iScience}, 29(2), 2026.

\bibitem{Alonso2024}
Albert Alonso and Julius~B Kirkegaard.
\newblock Learning optimal integration of spatial and temporal information in
  noisy chemotaxis.
\newblock {\em PNAS Nexus}, 3(7):pgae235, 07 2024.

\bibitem{Liang_2023}
Senwei Liang, Aditya~N Singh, Yuanran Zhu, David~T Limmer, and Chao Yang.
\newblock Probing reaction channels via reinforcement learning.
\newblock {\em Machine Learning: Science and Technology}, 4(4):045003, oct
  2023.

\bibitem{Jia}
Yanwei Jia and Xun~Yu Zhou.
\newblock Q-learning in continuous time.
\newblock {\em J. Mach. Learn. Res.}, 24(1), January 2023.

\bibitem{7hcf-p1yk}
Somnath Paramanick, Umashankar Pardhi, Harsh Soni, and Nitin Kumar.
\newblock Spontaneous emergence of run-and-tumble-like dynamics in a robotic
  analog of chlamydomonas: Experiment and theory.
\newblock {\em Phys. Rev. Lett.}, 135:168301, Oct 2025.

\bibitem{PhysRevLett.118.098101}
Zhaojun Li, Qiuxian Cai, Xuanqi Zhang, Guangwei Si, Qi~Ouyang, Chunxiong Luo,
  and Yuhai Tu.
\newblock Barrier crossing in escherichia coli chemotaxis.
\newblock {\em Phys. Rev. Lett.}, 118:098101, Feb 2017.

\bibitem{10.1073/pnas.1108608108}
Milena~D. Lazova, Tanvir Ahmed, Domenico Bellomo, Roman Stocker, and Thomas~S.
  Shimizu.
\newblock Response rescaling in bacterial chemotaxis.
\newblock {\em Proceedings of the National Academy of Sciences},
  108(33):13870--13875, 2011.

\bibitem{PhysRevLett.108.128101}
Xuejun Zhu, Guangwei Si, Nianpei Deng, Qi~Ouyang, Tailin Wu, Zhuoran He, Lili
  Jiang, Chunxiong Luo, and Yuhai Tu.
\newblock Frequency-dependent escherichia coli chemotaxis behavior.
\newblock {\em Phys. Rev. Lett.}, 108:128101, Mar 2012.

\bibitem{PhysRevE.104.054614}
Marcel Gerhard, Ashreya Jayaram, Andreas Fischer, and Thomas Speck.
\newblock Hunting active brownian particles: Learning optimal behavior.
\newblock {\em Phys. Rev. E}, 104:054614, Nov 2021.

\bibitem{Landin}
S.~Muiños-Landin, A.~Fischer, V.~Holubec, and F.~Cichos.
\newblock Reinforcement learning with artificial microswimmers.
\newblock {\em Science Robotics}, 6(52):eabd9285, 2021.

\bibitem{Wang_2021}
Qixuan Wang and Hao Wu.
\newblock Mathematical modeling of chemotaxis guided amoeboid cell swimming.
\newblock {\em Physical Biology}, 18(4):045001, may 2021.

\end{thebibliography}
\end{document}